\documentclass[aps,pra,twocolumn,showpacs,preprintnumbers,amsmath,amssymb,superscriptaddress]{revtex4-1}
\usepackage{dcolumn}
\usepackage{bm}
\usepackage{graphicx}
\usepackage{times}
\usepackage{epstopdf}
\usepackage{color}
\usepackage[subrefformat=parens,labelformat=parens]{subfig}
\usepackage[pdftex,colorlinks,bookmarks = true]{hyperref} 
\usepackage[capitalize]{cleveref} 

\begin{document}

\definecolor{Red}{rgb}{1,0,0}
\definecolor{Blu}{rgb}{0,0,01}
\definecolor{Green}{rgb}{0,1,0}
\definecolor{Purple}{rgb}{0.5,0,0.5}
\newcommand{\red}{\color{Red}}
\newcommand{\blu}{\color{Blu}}
\newcommand{\green}{\color{Green}}
\newcommand{\purple}{\color{Purple}}
\newcommand{\rvec}{\mathbf{r}}
\newcommand{\nvec}{\mathbf{n}}
\newcommand{\ev}[1]{\langle#1\rangle}
\newcommand{\abs}[1]{\left|#1\right|}
\newcommand{\uone}{\ensuremath{\mathrm{U}(1)} }
\newcommand{\sutwo}{\ensuremath{\mathrm{SU}(2)} }
\newcommand{\ztwo}{\ensuremath{\mathbb{Z}_2} }
\newcommand{\im}{\mathrm{i}}

\title{Immiscibile two-component Bose Einstein condensates beyond mean-field approximation: phase transitions and rotational response}
\author{Peder Notto Galteland}
\affiliation{Department of Physics, Norwegian University of Science and Technology, N-7491
Trondheim, Norway}
\author{Egor Babaev}
\affiliation{Department of Theoretical Physics, The Royal Institute of Technology, 10691 Stockholm, Sweden}
\author{Asle Sudb\o}
\affiliation{Department of Physics, Norwegian University of Science and Technology, N-7491
Trondheim, Norway}
\date{\today}

\begin{abstract}
We consider a two-component immiscible Bose-Einstein condensate with dominating intra-species
repulsive  density-density interactions. In the ground-state phase { of} such a system only
one condensates is present. This can be viewed as a spontaneous breakdown of \ztwo symmetry. We
study the phase diagram of the system at finite temperature beyond mean-field approximation. In the
absence of rotation, we show that the system undergoes a first order phase transition from this
ground state to a miscible two-component normal fluid as temperature is increased. In the presence
of rotation, the system features a competition between vortex-vortex interaction  and short
range density-density interactions. This leads to a rotation-driven ``mixing" phase transition in a
spatially inhomogeneous state with additional broken $\uone$ symmetry. Thermal fluctuations in this
state lead to nematic two-component sheets of vortex liquids. At sufficiently strong inter-component
interaction, we find that the superfluid and \ztwo phase transitions split. This results in the
formation of an intermediate state which breaks only \ztwo symmetry. It represents two phase
  separated normal fluids with density imbalance.
\end{abstract}
\maketitle

\section{Introduction}
\label{sec:intro}
Bose Einstein condensates (BECs) serve as highly useful synthetic model systems for a wide variety
of real condensed matter systems, due to their tunable interactions using magnetic and optical
Feshbach-resonances.  By creating mixtures of the same boson in different hyperfine states, one
effectively creates multicomponent condensates \cite{Wieman97, Modugno2002, Thalhammer2008,
McCarron2011}. Furthermore, by using crossed lasers, one may set up lattice model systems with a
vast combinations of intersite hopping matrix elements, as well as intrasite  interactions,
both intra- and interspecies\cite{Rolston1997, Mlynek1997, Jessen1998, Hansch1998, Verkerk1998,
Inguscio2008}. This means that these model systems, apart from being interesting in their own right,
emulate various aspects of a plethora of condensed matter systems of great current interest, such as
multicomponent  superconductors, Mott-insulators, and even topologically nontrivial band insulators.
The latter follows from the recent realization of synthetic spin-orbit couplings in such condensates
\cite{bec_soc1,bec_soc2,2013Natur.494...49G}. Of particular interest is the physics of these systems
in the strong coupling regime.

Spinor condensates with two components of the order parameter represent a first step away from
ordinary single-component condensates. This extension opens up a whole vista of physics which has no
counterpart compared {to} single-component condensates, due to the wide variety of interspecies
couplings that may be generated. Thus, these synthetic systems display physics which is beyond what is
ordinarily seen in condensed matter systems.

The parameter range {where} inter-component density-density interactions exceed intra-component
density-density interactions signals the onset of immiscibility, or phase separation, of the two
components. Numerical works solving the Gross-Pitaevskii ground-state equations have also found
interesting vortex lattices in this regime
\cite{Kasamatsu2003,tsubota3,Tojo2010,cipriani,PhysRevA.81.033629,mueller02,PhysRevA.90.043629,mottonen,battye2002stable}.
The effect of the repulsive inter-component interactions overpowering the intra-component
interactions causes the condensate to form intertwined sheets of vortices\cite{makoto2}. The condition for
immiscibility is readily realized experimentally, using magnetic and optical Feschbach
resonances\cite{Papp2008, Tojo2010}.

In this paper, we present results of large-scale Monte-Carlo simulations {of a two-component Bose
Einstein condensate   at finite temperature.} In a previous work, we have considered in detail
the effect of thermal fluctuations for the case where the inter-component density-density interaction is less than
or equal to the intra-component interaction \cite{Galteland_2015}. {In the present paper}, we focus on the regime of
density-density interactions where the inter-component interactions is larger than the intra-component
interactions. This regime is qualitatively different from the case in which the intra-component
interactions dominate.

{Previous works have studied the effect of an inter-component density-density interaction on
  the rotation-induced non-homogeneous ground states. These works were mostly limited to two spatial
  dimensions solving the Gross-Pitaevskii ground-state equations
  \cite{Kasamatsu2003,tsubota3,Tojo2010,cipriani,PhysRevA.81.033629,mueller02,PhysRevA.90.043629,mottonen,battye2002stable,makoto2}.
  although certain aspects of the three-dimensional case were also studied at mean-field
  level \cite{tsubota3,battye2002stable}.
 Here, we extend previous works in the immiscible regime to the case of finite temperatures and
 higher dimensions, including the full spectrum of density and phase-fluctuation of the condensate
 ordering fields.}

\section{Definitions}
\label{sec:model}

\subsection{Model}

We consider a general Ginzburg-Landau(GL) model of an $N$-component Bose-Einstein condensate,
which in the thermodynamic limit is defined as

\begin{equation} \mathcal{Z}=\int\prod_i^N\mathcal{D}\psi_i
\text{e}^{-\beta H},
\end{equation}
where
\begin{align}
  H=\int
  d^3r\Bigg[&\sum_{i=1}^N\sum_{\mu=1}^3\frac{\hbar^2}{2m_i}\left|(\partial_\mu-\mathrm{i}\frac{2\pi}{\Phi_0}A_\mu)\psi_i\right|^2\nonumber\\
                            &+\sum_i^N\alpha_i\abs{\psi_i}^2+
\sum_{i,j=1}^Ng_{ij}\abs{\psi_i}^2\abs{\psi_j}^2\Bigg]
  \label{eq:genH}
\end{align}
is the Hamiltonian. Here, the field $A_\mu$ formally appears as a non-fluctuating gauge-field
and parametrizes the angular velocity of the system. The fields $\psi_i$ are dimensionfull complex
fields, $i$ and $j$ are indices running from $1$ to $N$ denoting the component of the order parameter
(a ``color" index), $\alpha_i$ and $g_{ij}$ are Ginzburg-Landau parameters, $\Phi_0$ is the
coupling constant to the rotation-induced vector potential, and $m_i$ is the particle mass of species
$i$. For mixtures consisting of different atoms or different isotopes of one atom, the masses will
depend on the index $i$, while for mixtures consisting of same atoms in different hyperfine states,
the masses are equal among the components $i$.  The inter- and intra-component coupling parameters
$g_{ij}$ are related to real inter- and intra-component scattering lengths, $a_{ij}$, in the following
way
\begin{align}
  g_{ii} &= \frac{4\pi\hbar^2 a_{ii}}{m_i},\\
  g_{ij} &= \frac{2\pi\hbar^2 a_{ij}}{m_{ij}}; (i\neq j).
\end{align}
Here, $m_{ij}=m_i ~ m_j/(m_i+m_j)$ is the reduced mass. In this work, we focus on using BECs of homonuclear gases with
several components in different hyperfine states, i.e.  $m_i=m\,\forall\, i$. The system we primarily have in mind
is a mixture of Rb$^{87}$ atoms in two different hyperfine states $F=0$ and $F=1$, such that $N=2$. Note that when
$g_{12} = g_{21}\equiv \lambda g > g_{11}= g_{22} \equiv g$, i.e. $\lambda > 1$, there is a strong tendency in the
system to phase separate, leading to two immiscible quantum fluids. For a homonuclear binary mixture, such as the
mixture of Rb$^{87}$ atoms mentioned above, we have $m_{ij} = m_i/2$. Then, it suffices that $a_{ij} > a_{ii}$ for
the inter-component density-density interactions to dominate the intra-component {
density-density interactions}.

{In the following, we have introduced dimensionless coupling parameters and fields, following Appendix A of 
Ref. \onlinecite{Galteland_2015}.} It is convenient to rewrite the potential (repeated indices are summed over)
\begin{eqnarray}
V \equiv \alpha_i ~ |\psi_i|^2  +  g_{ij}\abs{\psi_i}^2 \abs{\psi_j}^2,
\label{potential-g}
\end{eqnarray}
by introducing interaction parameters $\eta, \omega$, such that $g = \eta + \omega$, and $\lambda g = \eta - \omega$,
i.e. $\eta=g(1+\lambda)/2$, $\omega=g(1-\lambda)/2$. Here, $\lambda$ denotes the ratio between the inter- and
intra-component interactions. Then, Eq. \ref{potential-g} takes the form (up to an additive constant)
\begin{eqnarray}
V & = & (\alpha_1+2 \eta) |\psi_1|^2 + (\alpha_2+2 \eta) |\psi_2|^2 \nonumber \\
& + & \eta(|\psi_1|^2+|\psi_2|^2-1)^2 + \omega(|\psi_1|^2-|\psi_2|^2)^2.
\label{potential-nw}
\end{eqnarray}
For $\lambda > 1$, $\omega < 0$, with the proviso that $\eta+\omega = \eta-|\omega| > 0$ for stability. Furthermore, we
will assume that $\alpha_1=\alpha_2$, such that $\langle |\psi_1|^2 \rangle = \langle |\psi_2|^2 \rangle$ when $\omega \geq 0$.
($\alpha_1 \neq \alpha_2$ would act as an external field conjugate to the pseudo-magnetization of the system, and would
destroy the Ising-like phase transition we report on below).

{Conversely, for a
binary mixture of homonuclear cold atoms,  one may express the ratios of intra- to inter-component scattering
lengths in terms of $\omega$ and $\eta$, as follows
\begin{eqnarray}
\frac{a_{12}}{a_{11}} = \frac{1-{\omega}/{\eta}}{1+ {\omega}/{\eta}}.
\label{ratio_a}
\end{eqnarray}
Note that the ratio of the scattering lengths only depend on the ratio  $\omega/\eta$.}

We discretize the model on a cubic lattice with sides $L$ by defining the order parameter field on a discrete
set of coordinates $\psi_i(\rvec)\rightarrow\psi_{\rvec, i}$, $\rvec \in (i \hat{\mathbf{x}}+j \hat{\mathbf{y}}+k
\hat{\mathbf{z}} |\, i,j,k = 1, \ldots, L)$. The covariant derivative is replaced by a forward difference,
\begin{equation}
D_\mu\psi_i(\mathbf{r})\rightarrow\frac{1}{a}\bigg(\psi_{\mathbf{r}+a\hat{\boldsymbol{\mu}},
i}\text{e}^{-\im aA_{\mu,\mathbf{r}}}-\psi_{\mathbf{r}, i}\bigg).
\end{equation}
{Here, the lattice version of the non-fluctuating gauge field is parametrized in Landau gauge, $A_{\mu, \rvec}=(0, 2\pi fx, 0)$, 
where $f$ is the number of vortices per plaquette, or filling fraction.} The lattice spacing, $a$, is fixed to be smaller than 
the characteristic length scale of the variations of the order parameter, and $\hat{\boldsymbol{\mu}}\in
(\hat{\mathbf{x}},\hat{\mathbf{y}},\hat{\mathbf{z}})$ is a unit vector.

Thus, the  lattice version of the Hamiltonian we consider is given by
\begin{align}
  H=-&\sum_{\substack{\rvec,\hat{\boldsymbol{\mu}}\\i}}\big|\psi_{\rvec+\hat{\boldsymbol{\mu}},
i}\big|\big|\psi_{\rvec,i}\big|\cos(\theta_{\rvec+\hat{\boldsymbol{\mu}},
i}-\theta_{\rvec,i}-A_{\mu,\rvec})\nonumber\\
+ & \sum_{\substack{\rvec,i}} (\alpha_i + 2 \eta) ~ \big|\psi_{\rvec,i}\big|^2\nonumber\\
+ &\sum_\rvec\eta{(\left|\psi_{\rvec,1}\right|^2+\left|\psi_{\rvec,2}\right|^2-1)}^2\nonumber\\
  +&\sum_\rvec\omega{(\left|\psi_{\rvec,1}\right|^2-\left|\psi_{\rvec,2}\right|^2)}^2.
\label{eq:ham}
\end{align}
Here, we have written the order parameter fields as real amplitudes and phases, $\psi_{\rvec,i} =
\left|\psi_{\rvec,i}\right|\text{e}^{i\theta_{\rvec, i}}$. {In addition, we have defined an
energy scale, $J_0 = \alpha_0^2a^3/g_0^3$, where $\alpha_0$ and $g_0$ are the parameters of the  Ginzburg-Landau
theory at $T=0$.} Throughout, we fix $\eta=5.0$ and $\alpha_1 + 2 \eta = \alpha_2 + 2 \eta = 0$.  
This guarantees a non-zero ground state condensate density for all values of $\eta$.

\subsection{Ground state symmetry}

Eq.~\ref{eq:ham} defines two superfluids coupled by density-density interactions. When there is no phase
separation,  we have a $\uone\times\uone$ symmetry broken in the ground state. When  the inter-component
interaction is equal to the intra-component interaction the system breaks $SU(2)$ symmetry. Here, we are
interested in the phase separated case. In this case, the system breaks an additional \ztwo symmetry, corresponding
to interchanging $\psi_1 \leftrightarrow \psi_2$.  That is, when $\omega > 0$, $|\psi_1|^2 = |\psi_2|^2$ is favored.
This represents a \ztwo-symmetric state. On the other hand, when $\omega < 0$,  $|\psi_1|^2 \neq |\psi_2|^2$  is
favored, such that $|\psi_1|^2 - |\psi_2|^2$ may acquire a nonzero expectation value, with equal probabilities
that the expectation value is either positive or negative. This corresponds to breaking an Ising-like $\ztwo$
symmetry. Thus, the ground state breaks a composite $U(1)\times Z_2$ symmetry.

\subsection{Observables}
\label{sec:Observables}
The equilibrium phases the model are characterized by several order parameters. To identify the Ising-like,
phase separated order of the system we define
\begin{equation}
\Delta = \abs{\ev{\abs{\psi_1}^2}-\ev{\abs{\psi_2}^2}},
\end{equation}
where $\ev{\abs{\psi_i}^2}$ is the thermal and spatial average of $\abs{\psi_i(\rvec)}^2$
\begin{equation}
  \ev{\abs{\psi_i}^2} = \ev{\frac{1}{L^3}\sum_\rvec\abs{\psi_{i,\rvec}}^2}.
\end{equation}
A finite value of $\Delta$ signals relative density depletion in either of the condensates. In addition to \ztwo
order, it is important to monitor the \uone ordering of the system. The helicity modulus measures
phase coherence along a given direction of the system. It is defined as
\begin{equation}
\Upsilon_{\mu,i}= \frac{1}{L^3}\frac{\partial^2 F[\theta^\prime]}{\partial\delta^2_\mu}\Bigg|_{\delta_\mu=0}.
\end{equation}
{Here, $F[\theta^\prime]$ is the free energy with an infinitesimal phase twist,
  $\delta_\mu$, applied along the $\mu$-direction,
\textit{i.e.}, we make the replacement
\begin{equation}
  \theta_{\rvec,i}\rightarrow\theta_{\rvec,i}^\prime = \theta_{\rvec,i} - \boldsymbol{\delta}\cdot{\bf
  r}
\end{equation}
in $F$.}

We also identify the nature of the phases by computing {thermal averages of real-space
configurations} of densities
$\ev{\abs{\psi_i(\rvec_\perp)}^2}$ and vortices $\ev{\abs{n_i(\rvec_\perp)}^2}$ in the system.
These are computed by averaging the quantity along the $z$-direction of the system, with subsequent thermal
averaging. That is,
\begin{equation}
  \ev{n_i(\rvec_\perp)} = \ev{\frac{1}{L_z}\sum_zn_{i,\rvec}}.
\end{equation}
and
\begin{equation}
  \ev{\abs{\psi_i(\rvec_\perp)}^2} = \ev{\frac{1}{L_z}\sum_z\abs{\psi_{i,\rvec}}^2}.
\end{equation}
The vorticity, $n_{i,\rvec}$ is calculated by traversing a plaquette with surface normal in the $z$-direction, adding
the phase difference $\theta_{\rvec+\hat{\boldsymbol{\mu}}, i}-\theta_{\rvec,i}-A_{\mu,\rvec}$ on each link. If this plaquette
sum turns out to  have a value outside  the primary interval, $(-\pi, \pi]$, $2 n \pi(-2 n \pi)$ is added to the sum, which
inserts a vortex of charge $+n(-n)$ on the plaquette.

To further characterize vortex structures, we examine the structure factor of the vortices, defined as
\begin{equation}
  S_i({\bf q}_\perp) =
  \Bigg\langle\abs{\frac{1}{L^3 f}\sum_{\rvec_\perp,z}n_{i,\rvec}e^{i{\bf
  q}_\perp\cdot\rvec_\perp}}^2\Bigg\rangle.
\end{equation}
This is simply the Fourier-transform of the $z$-averaged vorticity. To improve the resolution of the
interesting $q$-vectors, we remove the $\bf{q}_\perp=0$ point from the figures. We also compute the
specific heat capacity
\begin{equation}
  \frac{C_V L^3}{\beta^2} = \langle{(H - \langle H\rangle)}^2\rangle.
\end{equation}
as a means of precisely locating the various transition points.

\subsection{Details of the Monte-Carlo simulations}
\label{sec:MCdetails}
We consider the model on a lattice of size  $L_x \times L_y \times L_z$, using the Monte-Carlo
algorithm, with a simple restricted update scheme of each physical variable, and Metropolis-Hastings
\cite{Metropolis53,Hastings70} tests for acceptance. Here, $L_i$ is the linear extent of the system
in the Cartesian direction $i \in (x,y,z)$. In all our simulation, we have used cubic systems
$L_x=L_y=L_z=L$, with $L \in \{ 24, 32, 40, 48, 56, 64, 96, 128 \}$. At each inverse temperature,
$10^6$ Monte-Carlo steps are typically used, while $10^5$ additional sweeps are used for
equilibration. Each Monte-Carlo step consists of an attempt to update each amplitude and phase
separately in succession, at each lattice site. {To improve acceptance rates, we only allow
  each update to change a variable within a limited interval around the previous value, the size of which is
  chosen by approximately maximizing acceptance rates and minimizing autocorrelations.} The
  Mersenne-Twister algorithm is used to generate the pseudo-random numbers
  needed\cite{Matsumoto_MT}. To ensure that the state is properly equilibrated, time series of the
  internal energy measured during equilibration are examined for convergence. To avoid metastable
  states, we make sure that several simulations with identical parameters and different initial
  seeds of the random number generator anneals to the same state. {Measurements are
  post-processed with multiple histogram reweighting\cite{Ferrenberg_Swendsen}. Error estimates are
determined by the jackknife method\cite{Berg_jackknife}.}

\section{Phase diagram in the absence of rotation}
The model has a $\uone\times\uone\times\ztwo$ symmetry and hence the possibility for
several phase transitions. However,
for the parameter set which was simulated a single first-order phase transition is found, as we illustrate below
for $\omega = -2$. For $f=0$, we have considered $L\in\{24, 32, 40, 48, 56, 64\}$, and we present the results
for $L=64$.

\begin{figure}
\centering
\includegraphics[width=\columnwidth]{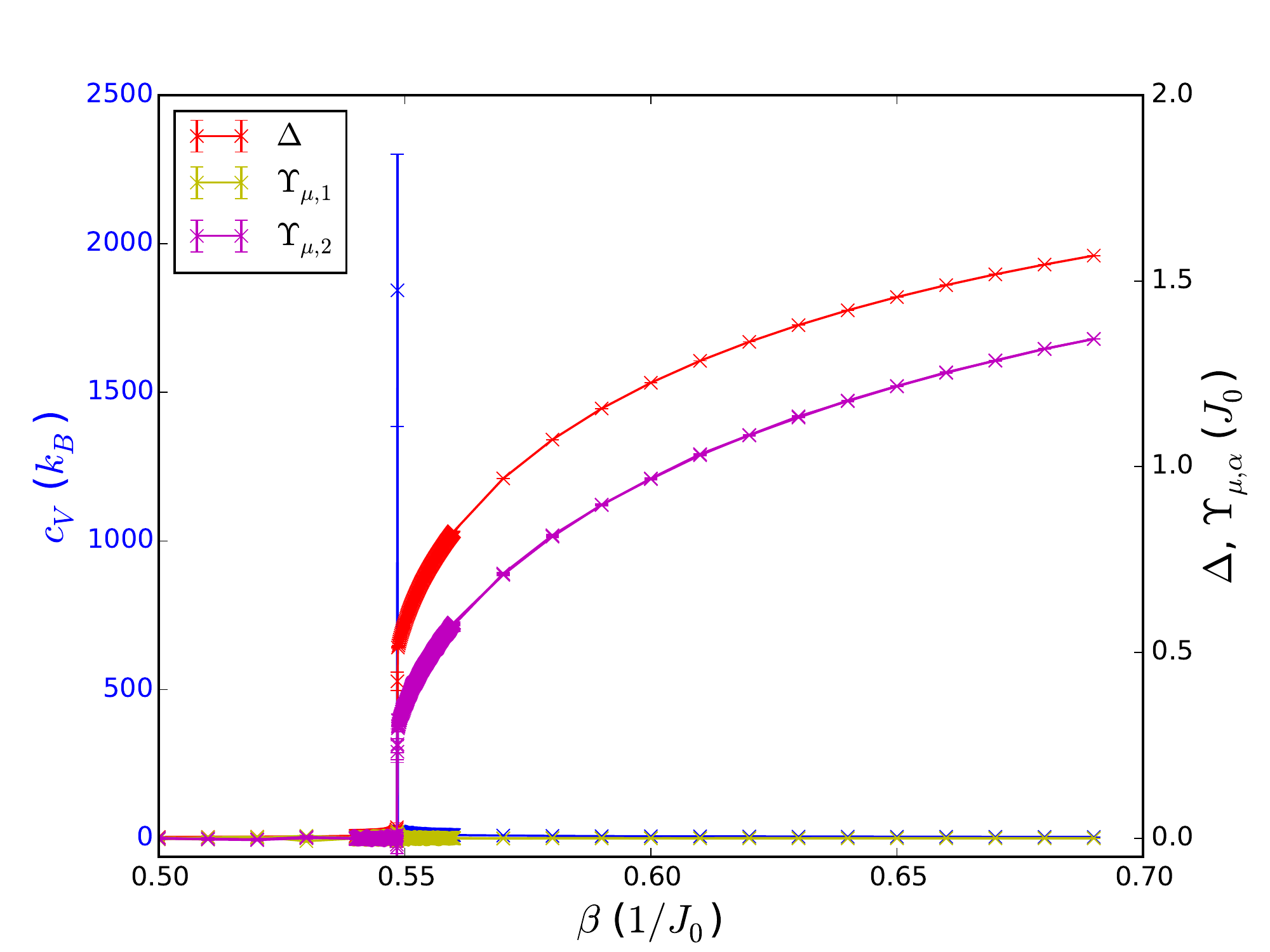}
\caption{The specific heat, $\Delta$, and helicity moduli $\Upsilon_{\mu,i}$ for $L=64$ for $\omega=-2$.
The specific heat features a $\delta$-function anomaly, characteristic of a first-order phase transition.
$\Delta$ has an onset at the same value that the specific heat anomaly is found. The helicity modulus of
the condensate with the smallest density is zero for all values of $\beta$. Note also the discontinuities
in the \ztwo-order parameter $\Delta$, and the \uone-order parameters $\Upsilon_{\mu,i}$. The origin of the
first-order character of the composite \ztwo $ \times $ \uone phase-transition is due to interaction between
domain walls and vortices. Details are explained in the text.
}
\label{cv_delta_hel}
\end{figure}

Fig. \ref{cv_delta_hel} shows the specific heat, Ising order parameter $\Delta$, and helicity moduli
(phase-stiffness or equivalently superfluid density) of both components for $L=64$ and $\omega=-2$.
The onset of the Ising-like order parameter $\Delta$ is located at the same temperature as the
$\delta$-function anomaly of the specific heat. Note that $\Delta$ vanishes {\it discontinuously} as
$\beta$ approaches the transition point from above.  Fig. \ref{cv_delta_hel} also shows the helicity
moduli of both components above and below the Ising transition.  In the high-temperature
\ztwo-symmetric phase, where both components have equal densities $\langle \abs{\psi_{i,\rvec}}^2
\rangle$, $\Upsilon_{\mu,i} = 0$. The onset of $\Upsilon_{\mu,i}$ signals a broken \uone symmetry.
Systems simultaneously breaking $U(1)$ and \ztwo symmetries, can have two independent phase
transitions. These are driven by proliferation of vortex loops for the \uone transition and
proliferation of domain walls for the \ztwo transitions.  For the system in question, these
transitions are not independent. Domain wall excitations interact with vortices, and therefore
proliferation of these topological defects are not independent processes. 
In our system the non-trivial interplay between the \uone- and \ztwo- sectors leads to a single  phase-transition. A different example of
system where interacting \ztwo and \uone sectors lead to a  nontrivial phase diagram,
is the case of phase-frustrated multiband superconductors \cite{Bojesen_2014,bojesen2013}.

Consider lowering the temperature from the fully symmetric phase where $\Delta = \Upsilon_{\mu,i} =
0$. The \ztwo-symmetry is  broken at a certain temperature such that $\Delta \neq 0$, i.e. {$\langle
\abs{\psi_1}^2 \rangle \neq \langle \abs{\psi_2}^2 \rangle$.  Thus, one component gets a reduced average 
density and one gets an enhanced average density. } These densities determine  the bare phase-stiffnesses 
in the problem. Thus, the component with the largest density effectively has less phase fluctuations than the 
other one. {Furthermore}, due to suppression of one of the components, the helicity modulus belonging to the 
dominant component becomes non-zero, while the helicity modulus belonging to the minor component can remain
zero. This effect is enhanced as temperature is lowered, since $\Delta$ increases monotonically
with $\beta$, thus decreasing bare phase stiffness of the minor component as $\beta$ is increased.
The low-temperature phase is therefore a two-component Bose fluid where one component is in a
\uone-ordered phase-coherent state and the other is in a \uone-disordered phase-incoherent state.
The spontaneous appearance of a disparity in densities reflects a broken \ztwo symmetry.

Breaking a \ztwo- or \uone-symmetry is usually associated with a second order phase transition,
i.e. a critical point. We next discuss how the above situation instead leads to a first-order phase
transition. Consider heating up the system from deep within the low-temperature phase, described in
the previous paragraph, until the system approaches the vicinity of the transition to the fully
symmetric phase. Formation of domain walls implies regions of suppressed density as well as imposes
cutoffs to vorticity, thus enhancing phase fluctuations. On the other hand, thermally-induced vortices
can decrease domain-wall tension. Thus, both the \ztwo-transition and \uone transitions can take place
preemptively, compared to single-component models. Under such circumstances two phase transitions can
merge into a single first order phase transition.

A similar mechanism for producing a single first-order phase transition by merging second-order phase
transitions, has previously been discussed in different systems with competing topological excitations,
such as the so-called $s+is$ superconducting states \cite{Bojesen_2014}, in $U(1)\times U(1)$ systems
with electromagnetic and drag couplings\cite{dahl2008preemptive,Herland2010}, and  as an interpretation
of observed first order phase transitions in multicomponent gauge theories
\cite{kuklov2006deconfined,kragset2006,2005cond.mat..1052K,2004PhRvL..92c0403K}. The phenomenon therefore appears quite generic in the physics of
multi-component condensates.

\begin{figure}
\centering
\includegraphics[width=\columnwidth]{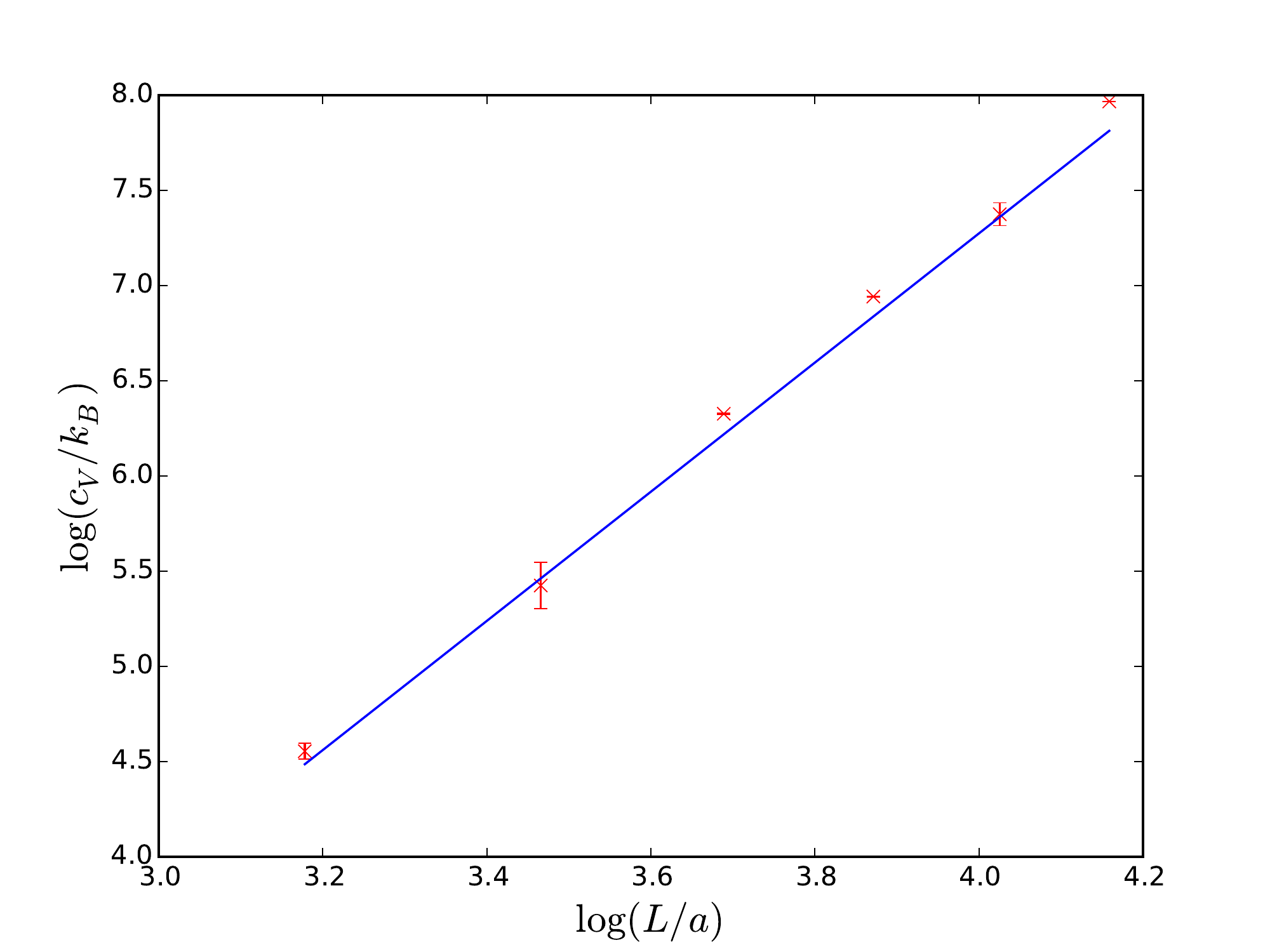}
\caption{Log-Log plot of the finite size scaling of the height of the specific
heat curves at $L\in\{24, 32, 48, 56, 64\}$. The parameters are $\omega=-2.0$,
$f=0$, and $\eta=5.0$. The exponent obtained is $\alpha/\nu=3.38(8)$.}
\label{fig:cvscale_f0}
\end{figure}

To corroborate the above discussion, we have performed finite size scaling of the specific heat peak
height for system sizes $L\in\{24, 32, 40, 48, 56, 64\}$. The scaling is shown in Fig.
\ref{fig:cvscale_f0} on a logarithmic scale, and the exponent obtained is $\alpha/\nu=3.38(8)$.
This is consistent with a first order transition, where the exponent would be $3$.

\begin{figure}
\centering
\includegraphics[width=\columnwidth]{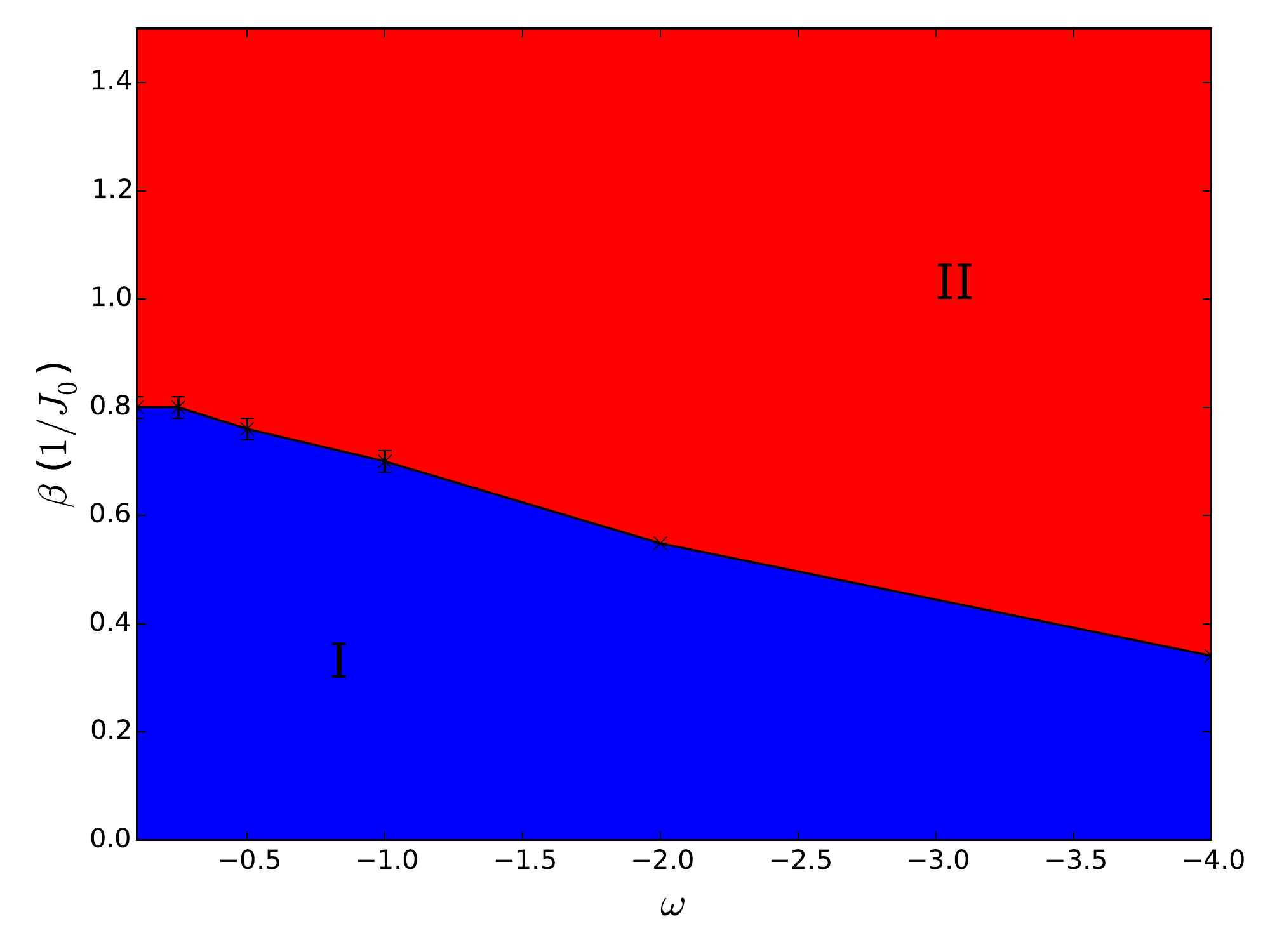}
\caption{
{The phase diagram of the two-component Bose Einstein condensate at zero rotation, $f = 0$, $\eta=5$, and $\omega < 0$.}
The high-temperature phase, region I, is a two-component miscible normal fluid. The low-temperature
phase, region II, is an immiscible (phase-separated) superfluid. The solid line separating these two
phases is a first-order phase transition where a composite \uone $\times$ \ztwo-symmetry is broken.}
\label{PD_f0}
\end{figure}

For $L=64$, we have performed similar computations for $w = \{-0.1, -0.25, -0.5, -1, -2 \}$.  The
results for $f=0$ are summarized in Fig. \ref{PD_f0}. The low-temperature phase is a two-component
immiscible (phase-separated) superfluid  with spontaneously broken \ztwo-symmetry, while the
high-temperature phase is a two-component miscible normal fluid. The solid line separating these two
phases is a first-order phase transition where a \uone- and a \ztwo-symmetry are broken simultaneously.
The endpoint of the phase-transition line at  $\omega=0$ is a point where the system acquires an
$SU(2)$-symmetry (see e.g.\cite{Galteland_2015}).

\section{Mixing and superfluid phase transitions in the presence of rotation}
We next consider the effect of imposing a finite rotation on the condensate. Our main results are presented
for a system size of $L=64$ and $f=1/32$, but we have considered system sized $L \in \{32,64,96,128 \}$.

Introducing a finite amount of (rotation-induced) vortices in the ground state significantly alters
the simple phase diagram presented in Fig. \ref{PD_f0}. The first effect is to suppress the
parameter regime where a broken \ztwo-symmetry is found, $\Delta \neq 0$.  Recall that for $f=0$,
any $\omega < 0$ sufficed to  bring about $\Delta \neq 0$ at sufficiently low $\beta$, as seen in
Fig. \ref{PD_f0}.  A finite amount of vortices alters this.
 Vortices interact via long range current-current interactions. It is energetically favorable
to maximize the distance between vortices, subject to the constraint that a specific number of them
has to be contained within a given area perpendicular to the direction of rotation. This effect
leads to a uniform distribution of minima (equivalently maxima) in the condensate densities.
On the other hand,
density suppression by vortices in one component in general allows the second to nucleate.
The short-range repulsive inter-component density-density interaction $(\eta -
\omega) (|\psi_1|^2 |\psi_2|^2 + |\psi_2|^2 |\psi_1|^2)$ (which exceeds the intra-component
density-density interaction $(\eta + \omega) (|\psi_1|^2 |\psi_1|^2 + |\psi_2|^2 |\psi_2|^2)$ for
$\omega < 0$), tends to produce regions where densities one component is large while the other is
small, and vice versa.
Below a critical value
of $-\omega =- \omega_c \approx +0.6$, we do not see any onset of $\Delta \neq 0$ at any value of $\beta$
as the system is cooled from a uniform state.
That is, the interface tension between the phases is sufficiently low
and the overall free energy, which includes long range inter vortex interaction
is minimized by  the state with $\Delta =0$.

\begin{figure}
\centering
\includegraphics[width=\columnwidth]{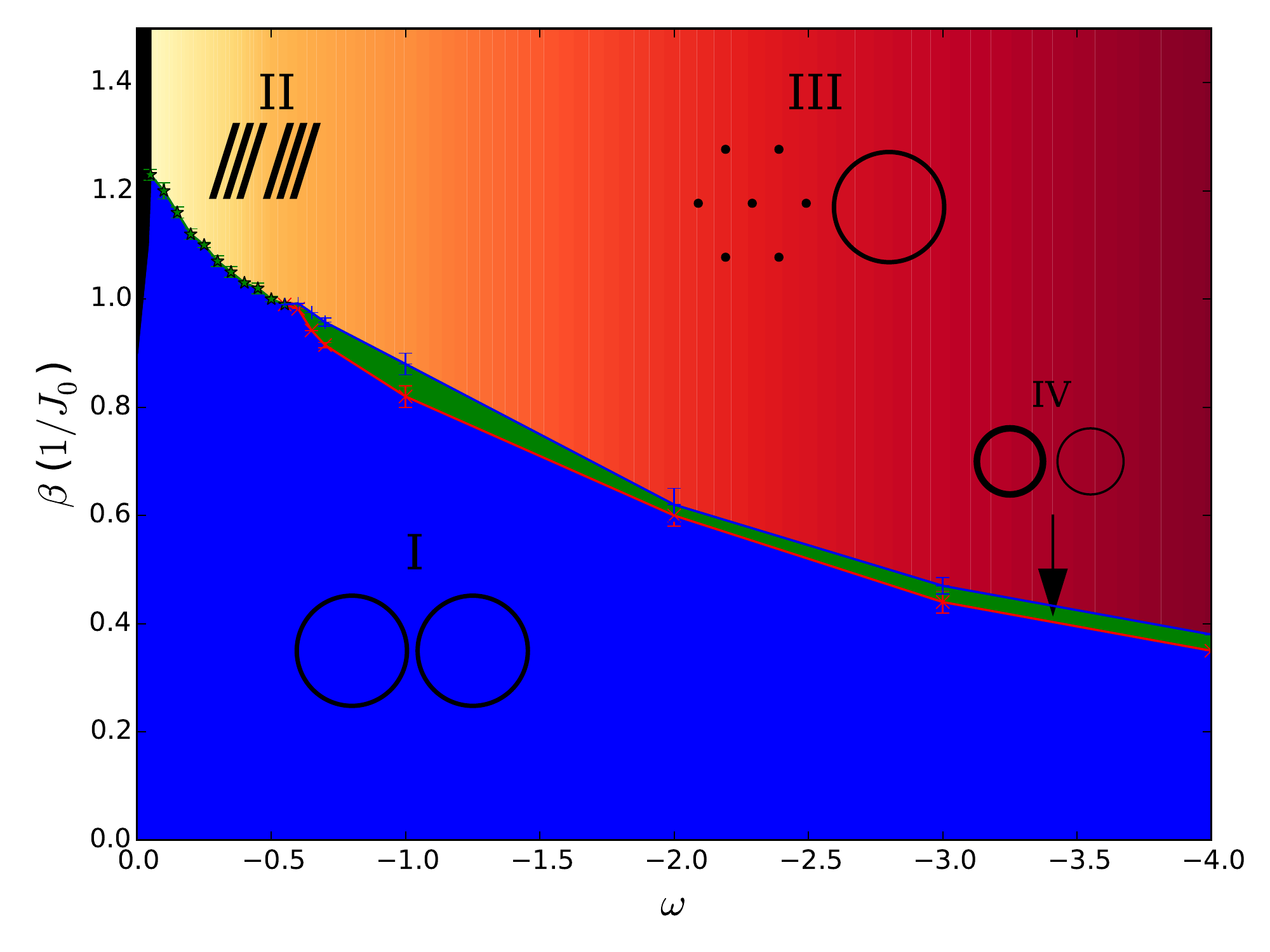}
\caption{{The  phase diagram of the two-component Bose Einstein condensate at finite rotation, $f \neq 0$, $\eta=5$,
and $\omega < 0$.} Negative $\omega$ may lead to the breaking of the \ztwo-symmetry in the problem, in addition to the
usual breaking of the obvious \uone $\times$ \uone - symmetry. Region I is a  \ztwo- as well as \uone $\times$ \uone-symmetric
two-component vortex-liquid phase. Region II is a \ztwo-symmetric striped (nematic) phase consisting of a two-component
vortex liquid with broken translational symmetry in a direction perpendicular to the stripes, but not in the direction
parallel to the stripes. Region III is a phase with broken \ztwo-symmetry, and with broken translational symmetry in one
condensate component, but not the other. Region IV is similar to Region III, except that no translational symmetry is
broken in either condensate component. Details are explained in the main body of the paper. }
\label{PD_f}
\end{figure}

For the subsequent discussion, it helps to consider a schematic phase diagram of the system with
$f \neq 0$, which we have obtained through large-scale Monte-Carlo simulations. The phase diagram is
shown in Fig. \ref{PD_f}. Region I denotes the simple translationally invariant high-temperature
\ztwo- and \uone $\times$ \uone-symmetric two-component phase with equal densities of both condensate
components. Region II shows the \ztwo-symmetric striped phase. Region III is a region with broken
\ztwo-symmetry, with one stiff  condensate component in a uniform hexagonal vortex lattice phase, and
one component in a uniform vortex liquid phase. Region IV is a region with broken \ztwo-symmetry, but
with the two condensates both in a vortex-liquid phase. Thus, the phase transition separating region
I from region II is a phase-transition line separating a two-component isotropic vortex liquid from
a two-component striped (nematic) vortex liquid. The line separating region I from Region IV is one
where a \ztwo-symmetry is broken, and the line separating region II from region IV is one where a
translational symmetry is broken and the system acquires
non-zero helicity modulus

\subsection{Transition from region I to region II}

We first consider the thermally driven transition from the high-temperature symmetric two-component
vortex liquid phase, region I, to the low-temperature two-component striped (nematic) phase, region
II, for fixed negative $\omega$, but where $|\omega| < |\omega_c|$, i.e. to the left of the
splitting point where Region IV opens up.

\begin{figure}
\includegraphics[width=\columnwidth]{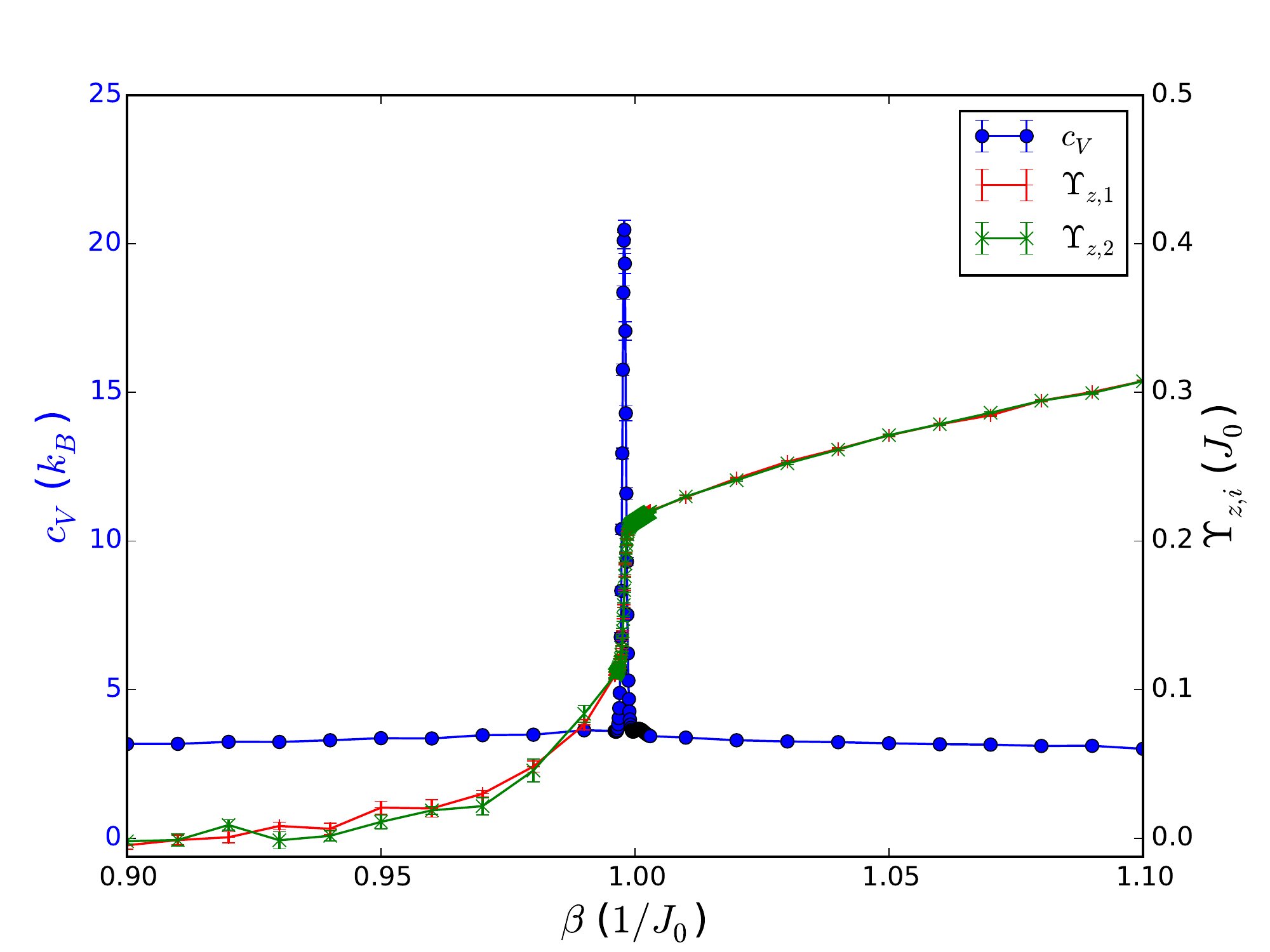}
\caption{Specific heat, $c_V$, and helicity moduli along the $z$-axis ,$\Upsilon_{z, i}$, with $f=1/32$ and $\omega=-0.50$,
i.e as the system transitions from Region I to Region II in Fig. \ref{PD_f}.}
\label{fig:u1u1_heat_heli}
\end{figure}

In Fig.~\ref{fig:u1u1_heat_heli} we show the specific heat $c_V$, helicity moduli in the $z$-direction $\Upsilon_{z, i}$ as
the inverse temperature $\beta$ is varied, for $f=1/32$ and $\omega=-0.50$. This corresponds to a value of $-\omega$ to the
left of the splitting point where Region IV opens up (see Fig. \ref{PD_f}). The longitudinal helicity moduli $\Upsilon_{z,i}$
of both components develop a finite expectation value. The onset of this finite value is accompanied by an anomaly in the
specific heat.

\begin{figure}
\includegraphics[width=\columnwidth]{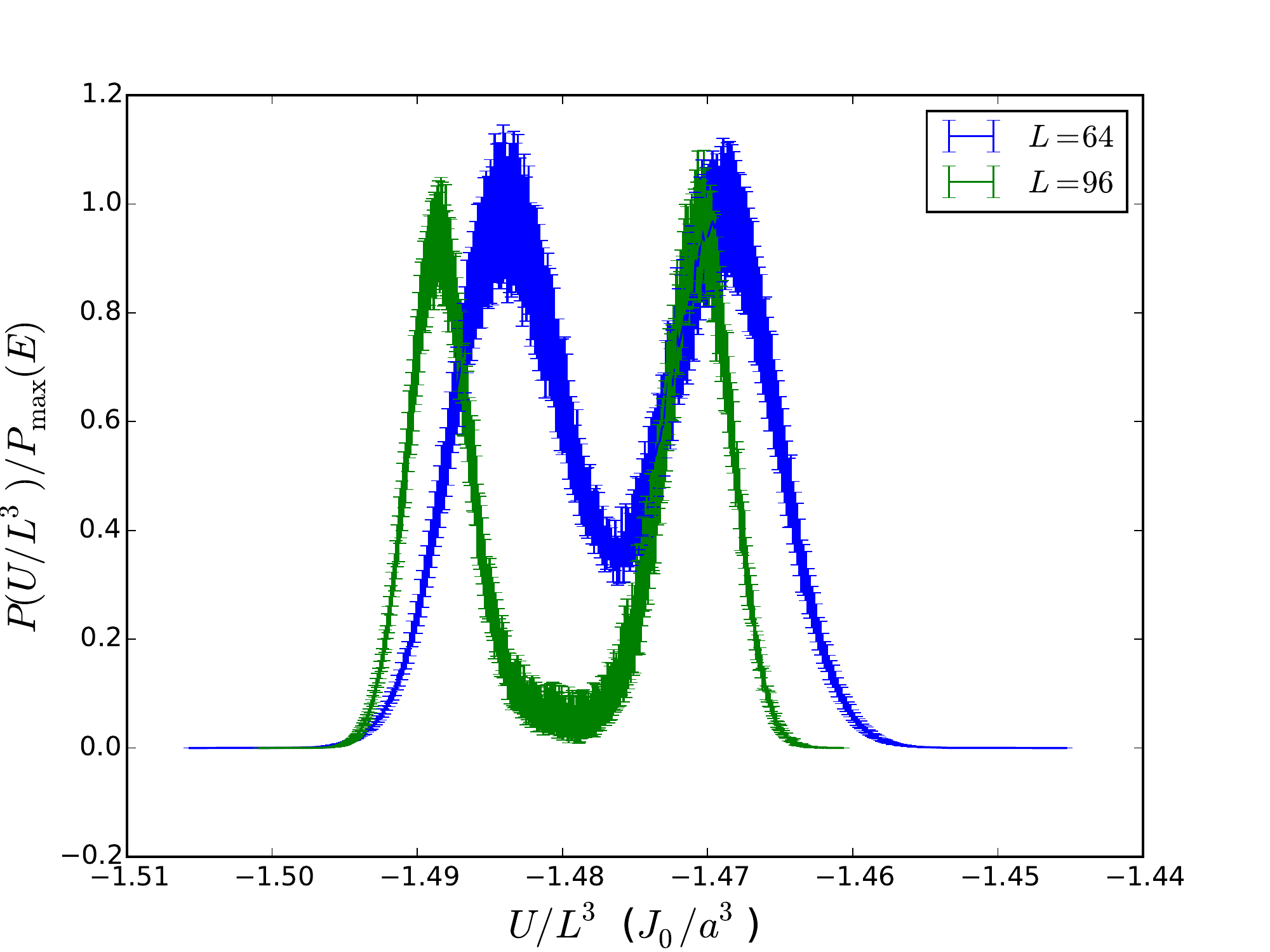}
\caption{Histograms of the probability distribution of the internal energy per site, $U/L^3$,  at the transition
point $\omega=-0.5, \beta\approx 0.9995$, separating Region I from Region II in Fig. \ref{PD_f}, for $L=\{64, 96\}$.
multi-histogram reweighting was used to obtain histograms with approximately equal peak heights. }
\label{fig:u1u1_double_peaks}
\end{figure}

We note the sharp, $\delta$-function anomaly in the specific heat and the discontinuous behavior of the helicity
moduli in both components. These features are  all straightforwardly interpreted as signals of a first-order phase transition.
This is furthermore borne out by performing a computation of the histogram of the free energy versus internal energy of the
system at precisely at the transition, see Fig. \ref{fig:u1u1_double_peaks}. This shows a double-dip structure with a peak in between,
the standard hallmark of two degenerate coexisting states separated by a surface whose energy is given by the height of the peak
between the minima. This surface energy clearly scales up with system size (more precisely it scales with the cross-sectional area
of the system), while the difference between the energies separating the two degenerate states approaches as finite value as
system size increases.  The histograms develop into two separate $\delta$-function peaks as the system size increases, while the
difference in the internal energy between the two degenerate states of equal probability (equivalently of equal free energy) is
the latent heat of the system. The latter clearly approaches a finite value per degree of freedom as the system size increases,
demonstrating the first-order character of the transition.

\begin{figure}
\centering
\subfloat{
\includegraphics[width=\columnwidth]{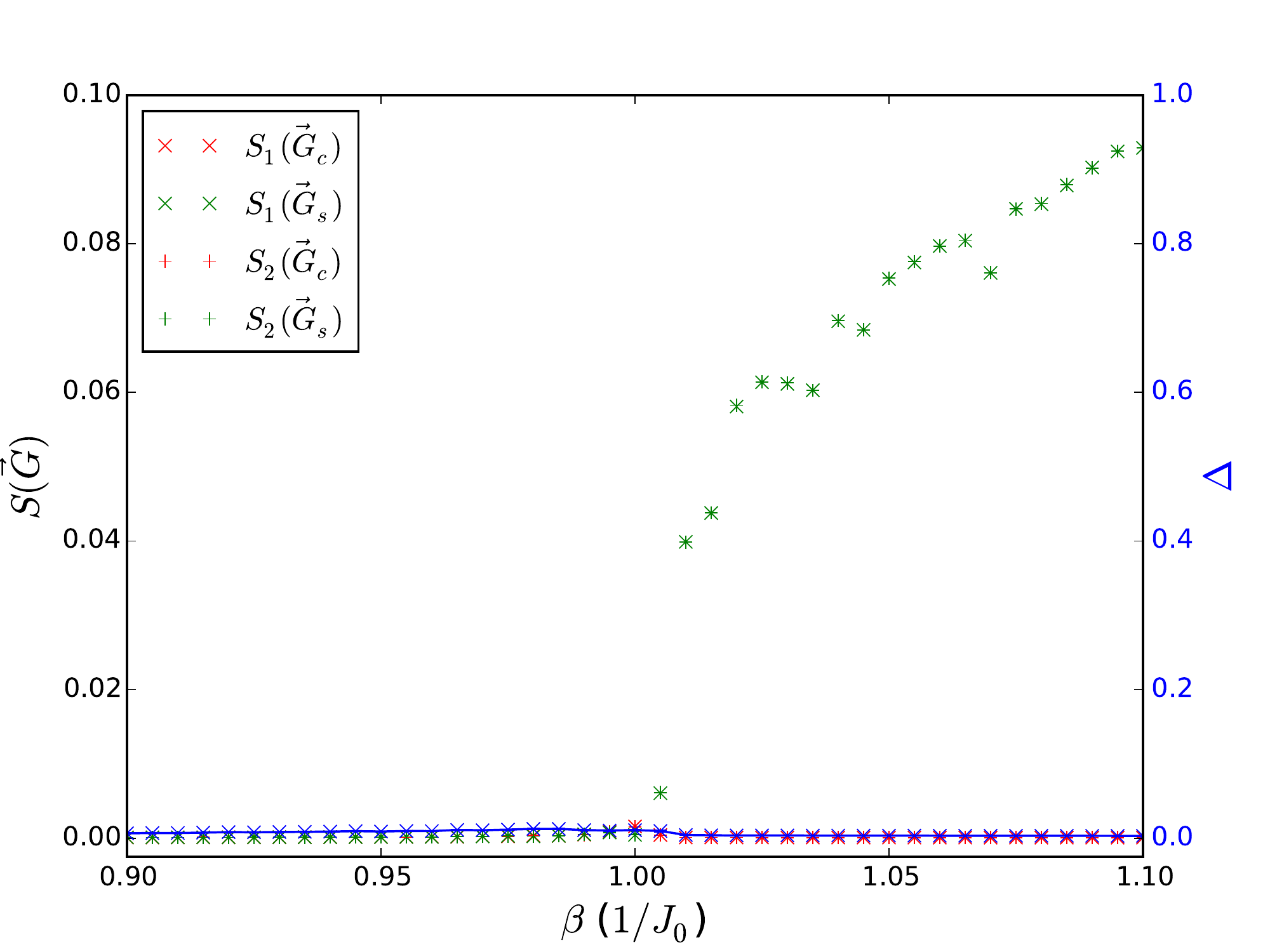}
\label{fig:u1u1_stripe_circle}}\\
\subfloat{
\includegraphics[width=\columnwidth]{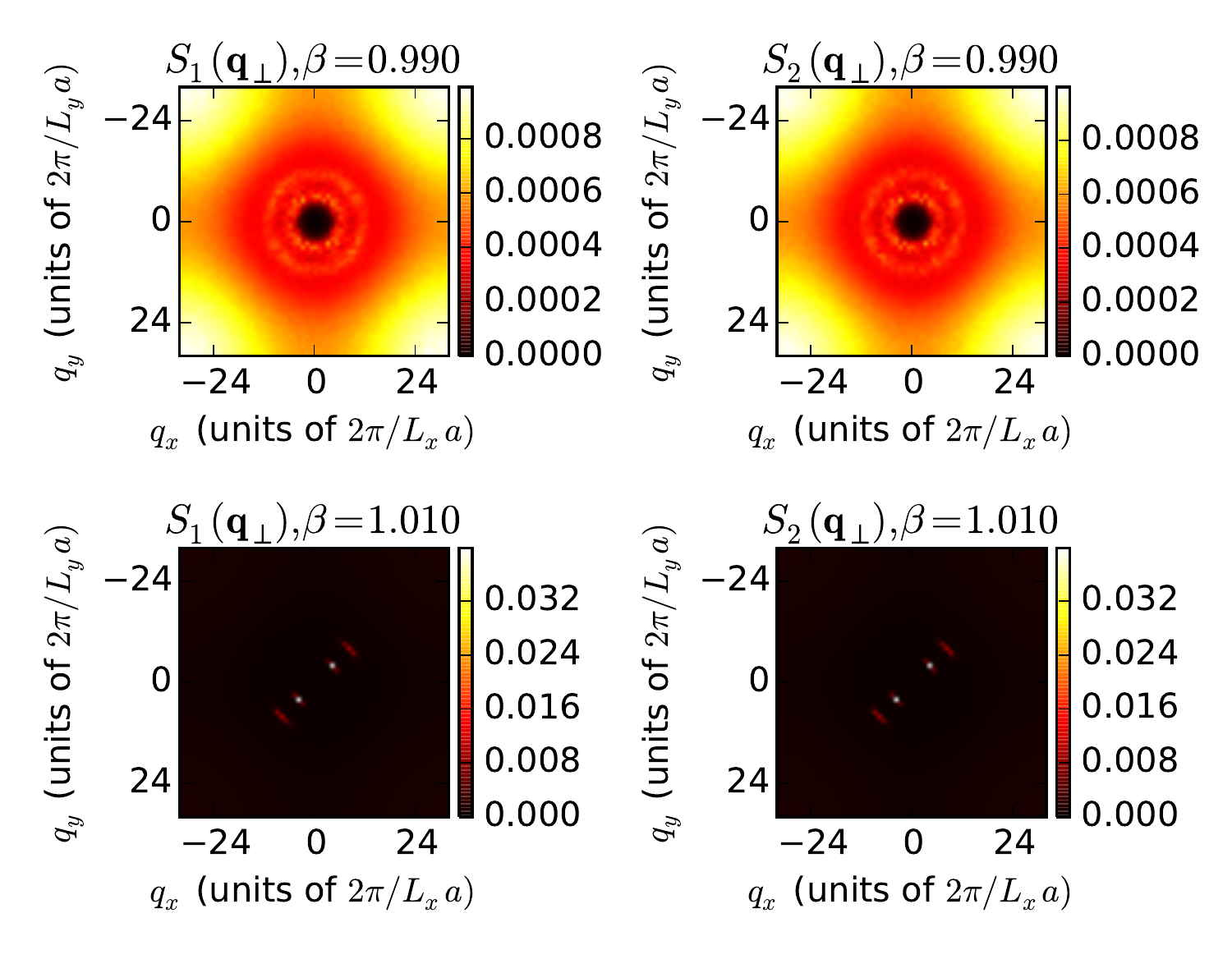}
\label{fig:u1u1_struct}}
\caption{\ztwo order parameter $\Delta$ and vortex structure functions $S_i({\bf q}_\perp), i \in  (1,2)$ in the vicinity of the
transition from Region I to Region II, Fig. \ref{PD_f}, with $f = 1/32,\omega=-0.50$. Panel (a) shows $\Delta$ as a function of
$\beta$, as well as structure functions at specific points in reciprocal space, $S_i({\bf G}_s)$ and $S_i({\bf G}_c)$. The
four bottom panels show the structure functions  $S_1 ({\bf q}_\perp)$  and  $S_2 ({\bf q}_\perp)$ for the two values
$\beta = 0.990$ and $\beta=1.010$.}
\label{u1u1_transition}
\end{figure}

To further characterize the transition  I $\rightarrow$ II, Fig. \ref{u1u1_transition}, shows the
\ztwo order parameter $\Delta$ and the structure functions $S_i({\bf q}_\perp), i \in (1,2)$ in a
narrow range around the transition point.  From the top panel, it is seen that $\Delta=0$ for all
$\beta$ considered. Moreover, we see that as $\beta$ is increased, the structure function evaluated
at a ${\bf q}$-vectors, ${\bf G}_c\equiv(\pm\pi/32, \pm\pi/32)$ on the Bragg-circle of the vortex-liquids are reduced, while the structure
function evaluated at Bragg peaks, ${\bf G}_s\equiv(\pm \pi/32, \mp \pi/32)$, corresponding to a uniform striped phase of $f=1/32$ increases.
The onset of the latter marks the transition from a uniform two-component vortex liquid to a
two-component nematic vortex liquid, a striped phase. The mechanism for producing the striped phase
is described  above. Note that in the thermodynamic limit, isolated vortex sheets can be expected to
be in the state of one dimensional liquid at any finite temperature in analogy with the absence of
crystalline order in one dimensional systems.

We thus conclude that the transition from Region I to region II is a first order phase-transition involving
the breaking of a composite \uone $\times$ \uone symmetry, from an isotropic two-component vortex liquid in
Region I to a two-component nematic phase of intercalated lattices of stripes of one-dimensional vortex liquids
in Region II. We next go on to consider in some more detail the structure functions, primarily to gain more
insight into the character of the striped phase of Region II.

The four bottom panels of Fig. \ref{u1u1_transition} show the structure functions $S_i({\bf q}_\perp), i \in (1,2)$
at two values of $\beta$, $\beta = 0.990$ and $\beta= 1.010$. At $\beta=0.990$, both structure functions show
ring-like structures characteristic of an isotropic liquid. Notice also that the intensity of the rings are
equal, which is a consequence of the fact that $\Delta =0$.  At  $\beta=1.010$, both structure functions have
developed Bragg peaks in one direction bot no Bragg peaks in the corresponding perpendicular direction. This is
indicative of a striped phase.

\begin{figure*}
\centering
\includegraphics[width=2\columnwidth]{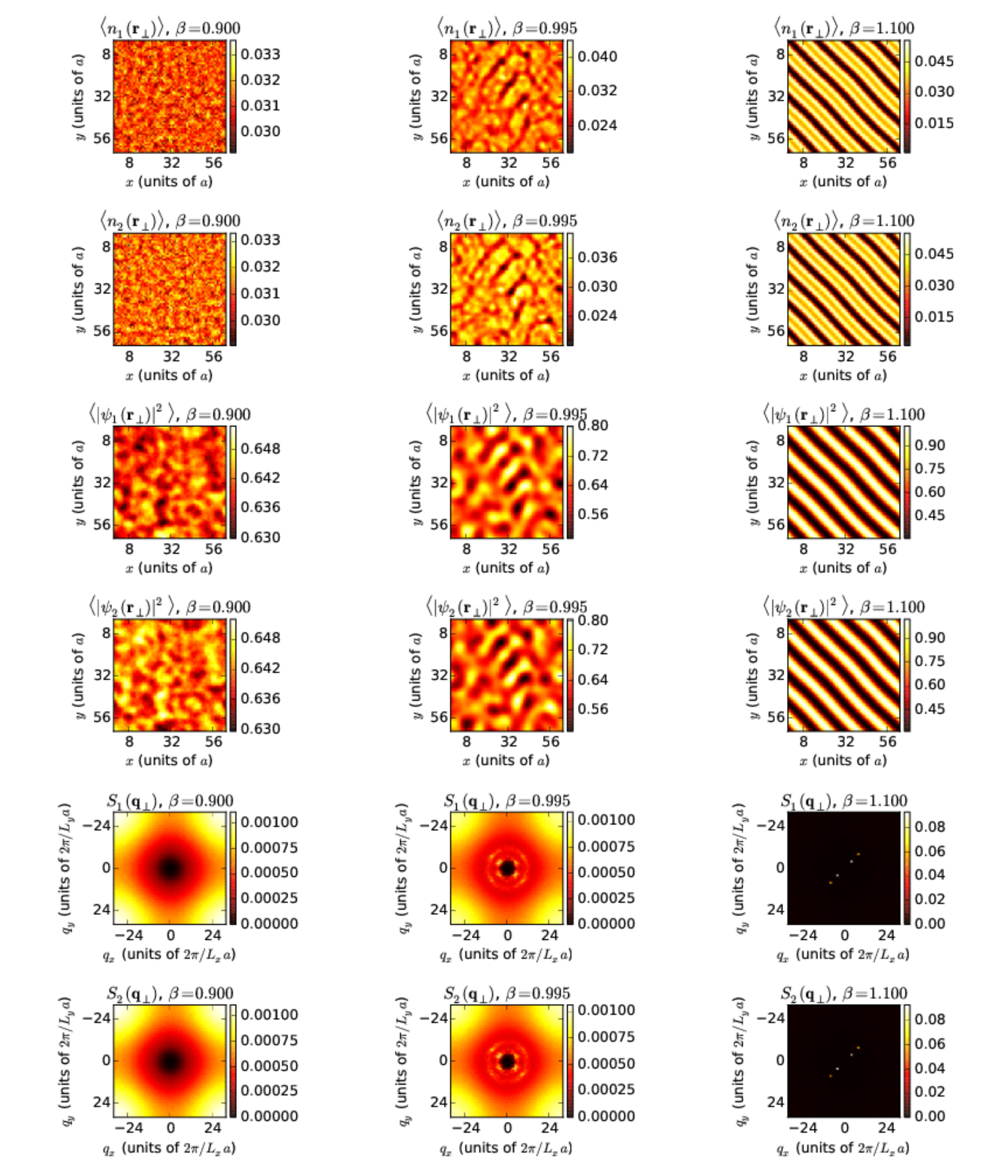}
\caption{Tableaux showing detailed real space and reciprocal space pictures of the transition from
region I to region II. The inverse temperature is varied between each column, $\beta\in\{0.900,
0.995, 1.100\}$. Each row show, in order, averaged vortex densities of each component,
$\ev{n_i(\rvec_\perp)}$, averaged amplitude densities of each component,
$\ev{\abs{\psi_i(\rvec_\perp)}}$, and vortex structure functions of each component, $S_i({\bf
q}_\perp)$. The first column corresponds to an inverse temperature well within region I, the second
column is at an inverse temperature just below where the transition into region II occurs, while the
last column is well within region II.}
\label{fig:tableaux_I_II}
\end{figure*}

This may be further corroborated by correlating the structure functions with real-space vortex structures for various values of
$\beta$. This is shown in \cref{fig:tableaux_I_II}

One aspect of the structure functions shown in the two bottom rows of \cref{fig:tableaux_I_II}, is
particularly important. Consider first the case $\beta = 0.900$, well within region $1$ for $\omega
< \omega_c$. This is shown in the first column of \cref{fig:tableaux_I_II}. The real-space vortex
configurations in both components are disordered. Moreover, $S_i({\bf q}_\perp), i \in (1,2)$ both
feature ring-structures characteristic of an isotropic liquid phase. The value of $|{\bf q}|$ at
which the rings appear is a measure of the average inverse separation between the vortices in the
isotropic liquids. The intensities of both structure functions is the same. Consider next the case
$\beta = 0.995$, shown in the second column of \cref{fig:tableaux_I_II}. From the real-space
pictures,  one discerns a tendency towards stripe-formation. This is reflected in  $S_i({\bf
q}_\perp), i \in (1,2)$, where the ring-like structures now instead are anisotropic, developing
peaks in the direction perpendicular to the direction of the incipient stripes. At even lower
temperatures $\beta = 1.100$, well within region II where stripes are fully developed, the tendency
towards anisotropies in $S_i({\bf q}_\perp), i \in (1,2)$ is even more obvious. This is shown in the
last column of \cref{fig:tableaux_I_II}. In this case, Bragg-peaks have fully developed in the
directions perpendicular to the stripes. There are, however, no Bragg peaks in the direction
parallel to the stripes. If the stripes were perfectly straight, there would be two weak Bragg peaks
in these directions.  This would be the one-dimensional analog of the ring-like liquid structures of
the isotropic liquid.  The value of $|{\bf q}|$ at which this single weak peak occurs corresponds to
the inverse average separation between vortices within the stripes. The reason they are not observed
in our calculations, is due to the slight fluctuations in the shape of the stripes, which wash the Bragg peaks out.

We thus conclude that region II is a striped phase where the stripes form one-dimensional (1D)
vortex liquids. Vortices in quasi-1D systems have finite energy and cannot form a 1D solid at any
finite temperature. This is consistent with the structure factor we observe. On the other hand, the
interaction between stripes may not be negligible, so the details of the phase diagram in Region II
warrant further investigation. A notable feature of this state is the finite helicity moduli in
$z$-direction, even if the structure factors show absence of vortex ordering within stripes. This
highly unusual situation originates with the positive interface energy between the two condensates.
That is, consider a stripe-liquid in $x$-direction. A vortex line in the $z$-direction is free to
execute transverse meanderings in the $x$-direction. A superflow in the $z$-direction would produce
a $y$-component of the Magnus-force on the $x$-components of the fluctuating vortex lines. However,
vortex segments are restrained from moving in $y$-direction due to the stripe interface tension.
This results in the observed finite helicity modulus in z-direction. Similar results are found for a
number of other $\omega$-values we have considered, for $-\omega < 0.6$. \footnote{We also
observed much smaller but finite helicity modulus in the direction perpendicular to stripes, which
we interpret as a consequence of weak standard geometric pinning of domain walls.}

\subsection{Transition from region I to region III, via region IV}

Increasing $-\omega$ further, such that the inter-species density-density interaction increases,
eventually favors a different pattern of phase-separation of the two components, despite the
uniforming effect of long-range current-current interactions between rotation-induced vortices. This
leads to a broken \ztwo-symmetry. The condensate component with a globally suppressed density will
therefore be in a vortex-liquid phase while the condensate component with globally enhanced density
will be in a vortex lattice phase. The combined preemptive \uone $\times$ \ztwo phase transition
found for $f=0$, now splits into two separate phase transitions. The splitting occurs because the
\uone-sector directly couples to the rotation, while the \ztwo-sector does not. The phase-transition
in the stiff \uone-sector, which is a vortex-lattice melting, is therefore separated from the
\ztwo-transition by an amount which depends on $f$.

Since $\Delta$ increases with $-\omega$ beyond $-\omega_c$, the transition temperature for the
\ztwo-transition increases. This effectively makes the dominant component stiffer as $-\omega$
increases. Thus, the temperature for melting the vortex lattice in the dominant \uone-sector also
increases with  increasing $-\omega$. The splitting between the \uone- and the \ztwo-sectors also
increases as $-\omega$ increases.

\begin{figure}
  \includegraphics[width=\columnwidth]{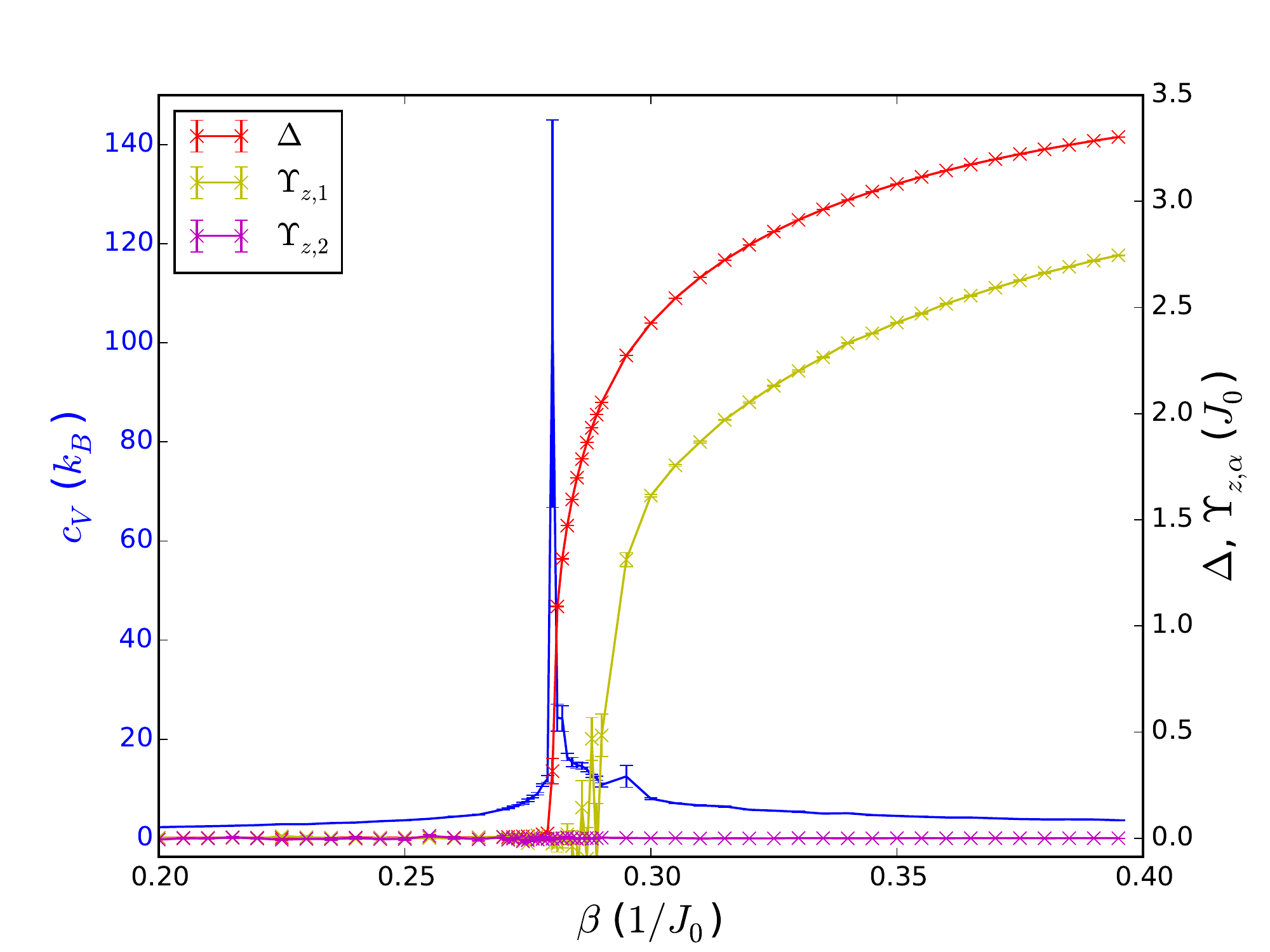}
  \caption{The phase transitions between Region I and Region IV, and between Region IV and Region III,
	  for $f=1/32, \omega=-4.0$, and $L=128$. Note the separation between the onset of $\Delta$ and
    $\Upsilon_{z,i}$.  The onset of $\Delta$ signals the breaking of a \ztwo-symmetry, along with
    the associated anomaly in specific heat $C_V$. This marks the transition from Region I to Region
    IV in Fig. \ref{PD_f}.  In Region IV, we have $\Delta \neq 0$, while both components remain in
    isotropic vortex liquid states.  In passing from Region IV to Region III in Fig. \ref{PD_f}, the
    onset of one of the helicity moduli, $\Upsilon_{z,1}$ say, signals the freezing of the vortex
    liquid in the corresponding component, while the absence of an onset of the helicity modulus,
    $\Upsilon_{z,2}$ say, in the other component shows that this component remains in a vortex
    liquid phase.  The onset of $\Upsilon_{z,1}$ signals the breaking of a \uone-symmetry associated
with vortex-liquid freezing. }
\label{fig:z2u1_delta_heli}
\end{figure}

\begin{figure}
\includegraphics[width=\columnwidth]{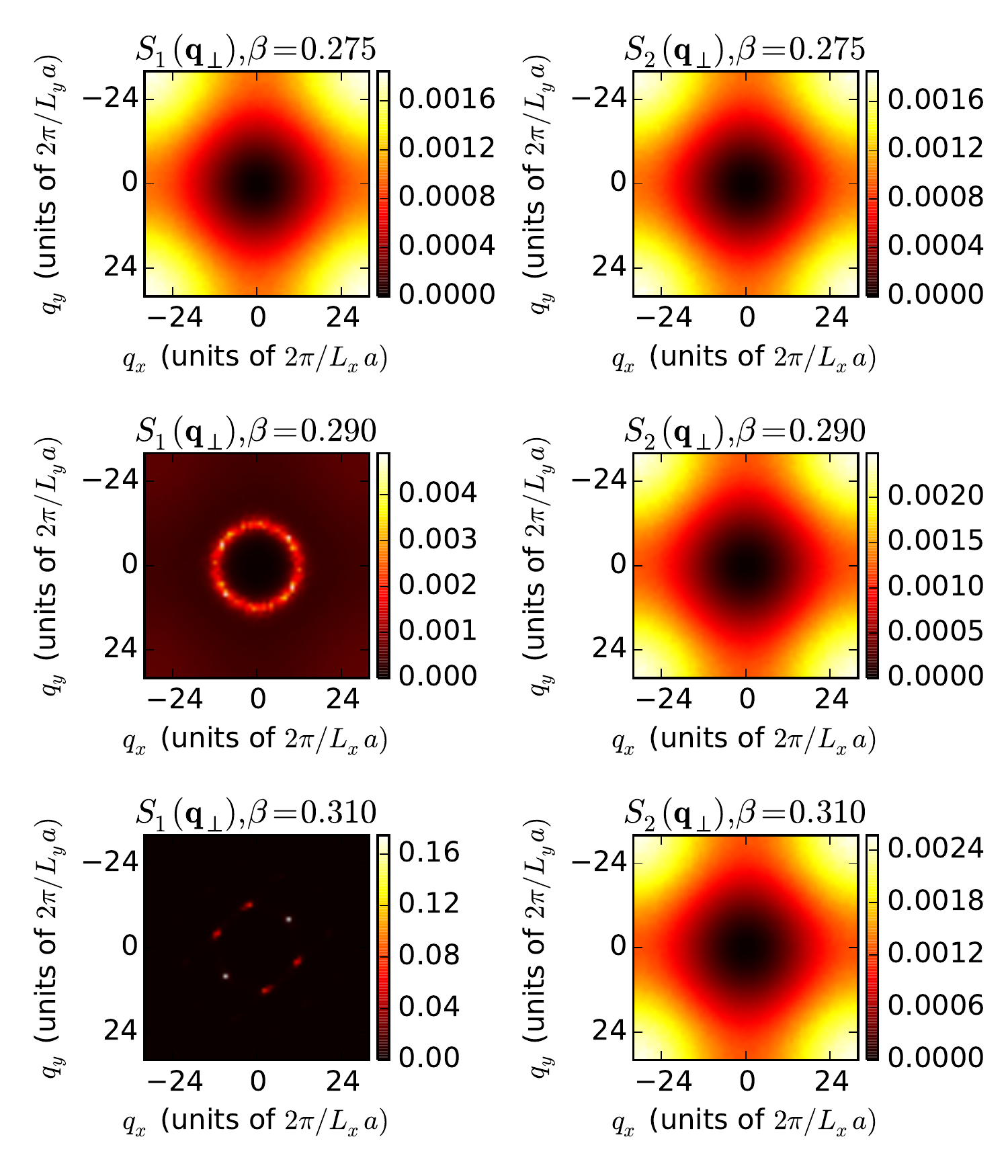}
\caption{The phase transitions of the system for $f=1/32, \omega=-4.0$. Structure functions $S_1 ({\bf q}_\perp)$ and
$S_2 ({\bf q}_\perp)$ at three different values of $\beta$, namely $\beta = (0.275, 0.290,0.310)$, corresponding
to Regions I, IV, and III in Fig. \ref{PD_f}, respectively.}
\label{fig:z2u1_struct}
\end{figure}

This is illustrated in Fig. \ref{fig:z2u1_delta_heli}, showing $\Delta$, specific heat $C_V$, and
$\Upsilon_{z,1}, \Upsilon_{z,2}$ as functions of $\beta$. The \ztwo order parameter $\Delta$ has an
onset at $\beta_{\ztwo}$, at which the specific heat has an anomaly.  There is no onset of
$\Upsilon_{z,1}$, showing that component $1$ remains in a vortex liquid phase. Component $2$
forms a vortex solid at lower temperature, as evidenced by the onset of $\Upsilon_{z,2}$. This
happens at a $\beta_{\uone}$ which is separated from $\beta_{\ztwo}$, as explained above.

Fig. \ref{fig:z2u1_struct} shows the structure functions $S_1 ({\bf q}_\perp)$ and $S_2 ({\bf
q}_\perp)$ at $\omega=-4.0$ at three different values of  $\beta$, namely $\beta = (0.275,
0.290,0.310)$. These values correspond to Regions I, IV, and III in Fig. \ref{PD_f}, respectively.
Here again, we see the freezing of one component across the transition, while the other component
remains in the liquid phase. The additional information we get out of these panels is that one
component remains an {\it isotropic} vortex liquid, while the other component freezes into a
hexagonal vortex liquid. This sets the low-temperature Region III (see Fig. \ref{PD_f}) at $\omega =
-4.0$ drastically apart from the low-temperature Region II (see Fig. \ref{PD_f}) at $\omega =
-0.50$. The latter features a low-temperature two-component nematic vortex liquid phase with broken
rotational invariance, the former case features a low-temperature mixed isotropic vortex
liquid/hexagonal vortex lattice phase.

\begin{figure*}
\centering
\includegraphics[width=2\columnwidth]{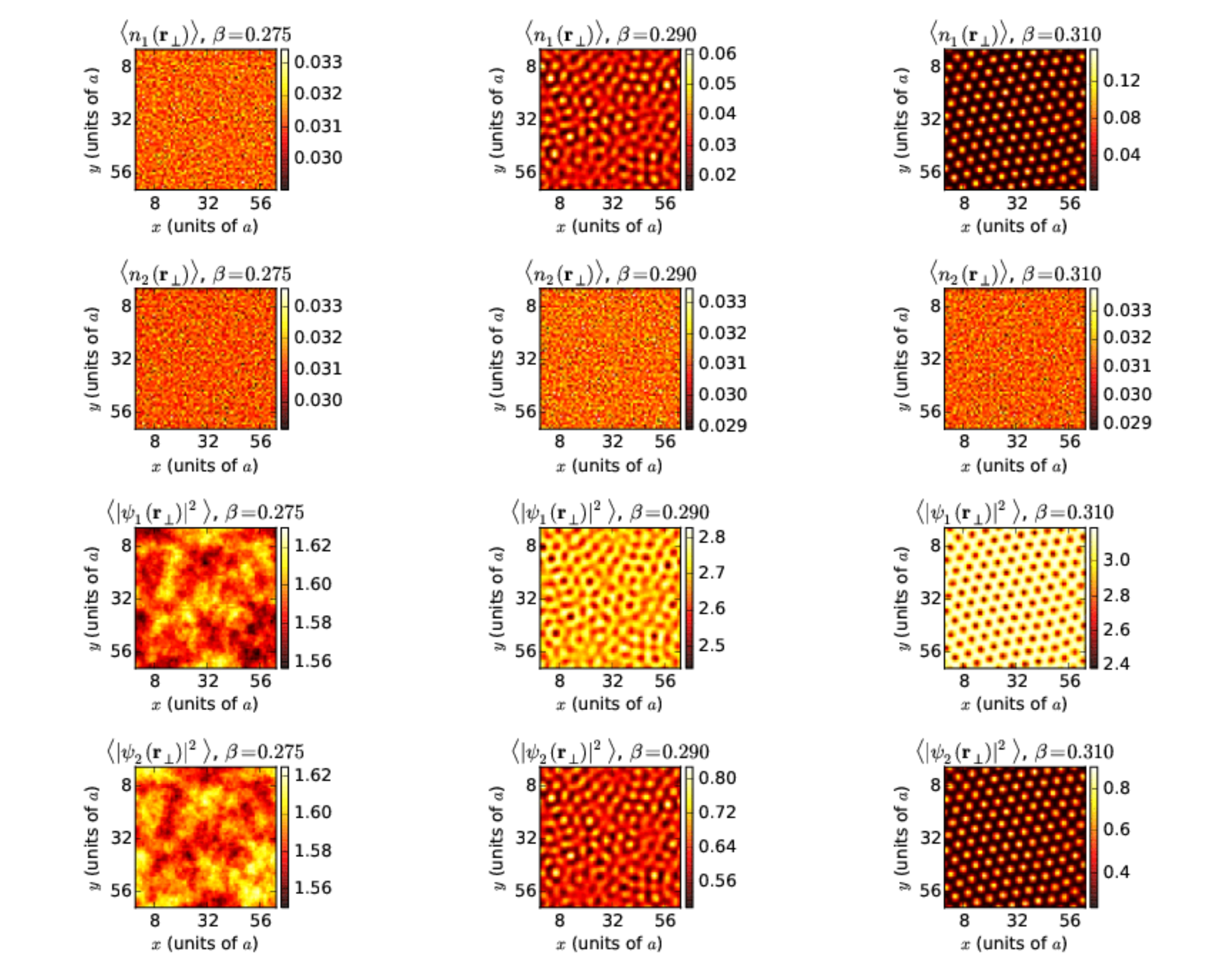}
\caption{Tableaux showing detailed real space pictures of the transition from region I to region III, via Region IV.
The inverse temperature is varied between each column, $\beta\in\{0.275, 0.290, 0.310\}$, and $\omega =-4$. Each row
shows, from top to bottom, averaged vortex densities of each component, $\ev{n_i(\rvec_\perp)}$, and averaged amplitude
densities of each component, $\ev{\abs{\psi_i(\rvec_\perp)}}$. Note the vortex-ordering in one of the components, and the lack
of vortex-ordering in the other component, as the system transitions from the symmetric phase region I ($\beta = 0.275$) to
the low-temperature phase region III  ($\beta = 0.310$). Note also the disparity in density-amplitudes in the two
components in the intermediate regime region IV ($\beta = 0.290$), due to the \ztwo-symmetry breaking. }
\label{fig:tableaux_I_IV_III}
\end{figure*}

For a more detailed overview of the transition, \cref{fig:tableaux_I_IV_III} show the evolution of
$\ev{n_i(\rvec_\perp)}$, $\ev{\abs{\psi_i(\rvec_\perp)}}$, and $S_i({\bf q}_\perp)$ across the three
regions, I, IV, and III. If one follows the evolution of the vortex densities in each component, it
is seen that the component which acquires a low stiffness in region IV and III is virtually
unchanged, \textit{i.e.} it remains in a completely uniform state. The other component, on the other
hand, evolves from a uniform state in region I, through being close to freezing into a hexagonal
lattice in region IV, and finally into a hexagonal structure in region III. The amplitude densities
corroborate this picture. In region I they are on average equal and uniform, while in region IV the
difference in stiffness is clearly seen. Here some inhomogeneities arise in the stiff component as
the vortices are close to entering a hexagonal phase, which is also reflected in the soft component
simply because of the local intercomponent repulsion. In region III, the amplitude density of the
stiff component is high and uniform with small dips corresponding to each vortex. The soft
component is low and uniform with small peaks, again due to intercomponent interactions.

\subsection{Transition from region II to region III}

\begin{figure*}
\centering
\includegraphics[width=2\columnwidth]{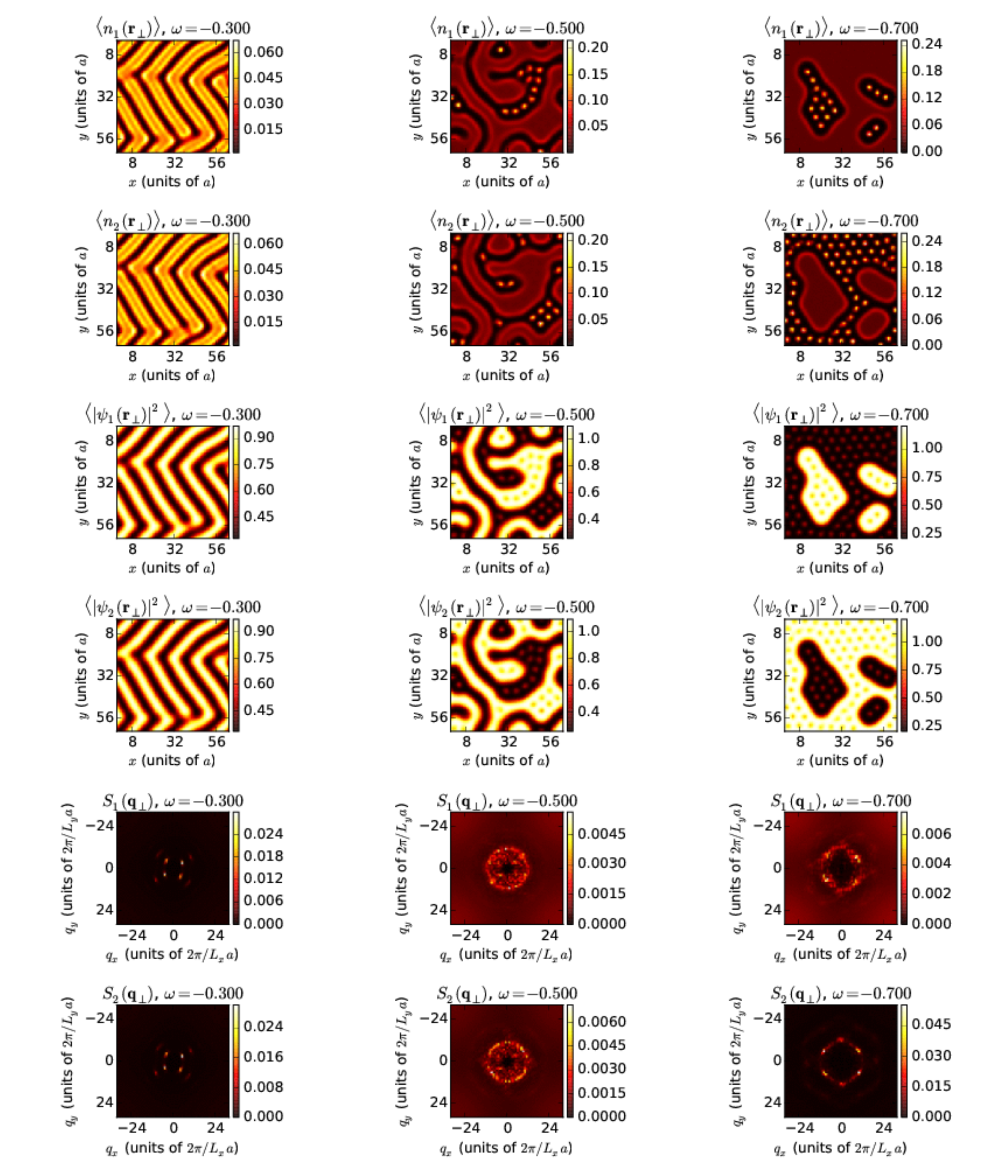}
\caption{Tableaux showing detailed real space and reciprocal space pictures of the transition from
  region II to region III. The parameter $\omega$ is varied between each column, $\omega\in\{-0.300,
-0.500, -0.700\}$. Each row shows, from top to bottom, averaged vortex densities of each component,
$\ev{n_i(\rvec_\perp)}$, averaged amplitude densities of each component,
$\ev{\abs{\psi_i(\rvec_\perp)}}$, and vortex structure functions of each component, $S_i({\bf
q}_\perp)$. The first column shows a configuration close to region II, the second column is a
configuration from the highly metastable crossover region, while the last column shows a
configuration close to region III.}
\label{fig:tableaux_II_III}
\end{figure*}

Finally, we consider the transition from Region II to Region III. In Region II, we have $\Delta = 0$, while in Region III, $\Delta \neq 0$.
Therefore the Regions II and III are  separated by a \ztwo-symmetry breaking. Stripe-forming systems in general have complicated structural
transitions. We find an intermediate regime where the lattice of stripes has disordered, but where the hexagonal lattice/isotropic
liquid-mixture has not yet fully developed. This results in multiple metastable, but robust coexisting phases of vortices in components 1
and 2 residing in different parts of the condensate. These  two coexisting phases are separated by a surface of positive surface energy. This
surface constrains the motion of vortex systems. As a result, in the finite systems which we simulate, the helicity moduli
$\Upsilon_{z,i}$ acquire nonzero values in both components in this intermediate regime.

As $-\omega$ is increased further, such that one component becomes dominant and the other is suppressed, the minor component becomes normal.
Note that when the inclusions of the normal component become isolated, they represent quasi-1D subsystems. Quasi-1D systems are superfluid
only at zero temperature. However, simulations on finite systems may still display finite helicity modulus.  As the density of the component
increases, the corresponding intra-component current-current interaction between the rotation-induced vortices in this component increases.
Hence, the intra-component long-range interaction for this component dominates, and a hexagonal vortex-lattice results. Consequently, the
helicity-moduli in the two components have quite different behavior as $-\omega$ increases. In the component that eventually takes up a
vortex lattice state, it increases monotonically with $-\omega$. In the other component, it is non-monotonic as a function of $\omega$,
eventually approaching $0$ deep into Region III.

Typical examples of the vortex structures that appear between Region II and Region III in Fig. \ref{PD_f} are shown in \cref{fig:tableaux_II_III}.
These are all metastable, long-lived states which prevent equilibration of the system. We have been unable to locate the sharp separatrix between
these two regions, and whether there are other stable intermediate phases due to the lack of equilibration.  Note that this  problem is known in
other stripe-forming systems where phases are separated by metastable and glassy states \cite{sellin2,julien}.

\section{Conclusions}
\label{sec:conclusions}
In this paper, we have considered the states of a two-component Bose Einstein  condensate in the situation where inter-component
density-density interactions  dominate over the intra-component density-density interactions. The two components of the condensate
are assumed to be comprised of homonuclear atoms in two different hyperfine states. The problem  features an Ising-like symmetry.
This Ising (or \ztwo) symmetry emerges from the dominance of the inter-component interactions over the intra-component ones. The spontaneous
breaking of this Ising-symmetry corresponds to a spontaneously generated, interaction-driven, imbalance between condensates in different
hyperfine states.

At finite rotation, we find four regions, denoted Regions I, II, III, and IV, of thermodynamically stable states, see Fig. \ref{PD_f}.
Region I is a high-temperature regime where the system remains in a two-component isotropic vortex liquid phase with equal densities of both
components, i.e. no  imbalance between condensate components in different hyperfine states. Region II is a nematic phase (broken
rotational symmetry) with ordered stripes of one-dimensional vortex liquids, and with no imbalance between components of different
hyperfine states. This state features a spontaneously broken composite \uone $\times $ \uone-symmetry, but is \ztwo-symmetric.
In addition it spontaneously breaks translation symmetry in one direction due to formation of periodic modulation of condensates.
Region III is a mixed state with one component in a \uone-symmetric isotropic vortex liquid phase while the other component resides in a
hexagonal vortex lattice phase with broken \uone-symmetry. The origin of the different behaviors of the two components is that Region
III also features a spontaneously broken \ztwo-symmetry, i.e. an imbalance between the density of one hyperfine state and the other.
The component with a large density has higher phase stiffness than the component with the lower density, hence the discrepancy in
their vortex states. Finally, Region IV is a region intermediate between Region I and Region III, in which \uone-symmetry is not
broken in either of the components, but where a spontaneously generated imbalance between densities of hyperfine states exists.
Both components are in an isotropic and disordered vortex state.

The phase transition from Region I to Region II in Fig. \ref{PD_f} is a first-order composite \uone $\times$ \uone transition.
The phase transition between Region I and Region IV is associated with a spontaneous \ztwo symmetry breaking where a density-imbalance
between condensates of different hyperfine states sets in. The phase-transition between Region IV and Region III is a first order
\uone transition associated with the freezing of an isotropic vortex liquid in one component into a hexagonal vortex lattice in the
same component, while the other component (the one with depleted density due to the \ztwo-symmetry breaking) remains in the isotropic
vortex liquid phase. The phase transition from Region II to Region III, driven by increasing the dominance of inter-component
density-density interactions over intra-component density-density interactions, involves at the very least a spontaneous breaking
of a \ztwo-symmetry as the two condensate
components pass from a nematic state of intercalated lattices of one dimensional vortex liquids into a mixed state of an isotropic
vortex liquid in one component and a hexagonal vortex lattice in the other component. Other than that, this transition is characterized by a
broad regime of metastable states with inhomogeneous phase separation.

{Fig. \ref{fig:tableaux_I_IV_III}  suggests that the rotation frequency is much smaller than the second critical frequency $\Omega_{c2}$. 
A very rough esitimate, based on core size, gives  $\Omega \propto 0.1\Omega_{c2}$. This puts the system well outside the regime of
lowest-Landau level physics. The system is therefore indeed in a regime where it makes sense to talk about vortex-degrees of freedom rather 
than zeroes of the order parameter as the relevant degrees of freedom. For this rotation frequency, we have found the critical value of 
$\omega$ (one of our interaction parameters) to observe phase IV to be $\omega_c \approx -0.6$. From this, we find from Eq. \ref{ratio_a} that
this requires scattering lengths $a_{12}/a_{11} > 1.3$. Since these scattering lengths {\it a priori} are very similar, and can be manipulated
with Feshbach resonances, it seems feasible to be able to observe phase IV. In order to see the striped ground states phase II, the
requirement is only that $a_{12}/a_{11}> 1$, which certainly seems to be within the realms of possibility.}

\begin{acknowledgments}
P.~N.~G. was supported by NTNU and the Research Council of Norway . E.~B. was supported by the Knut and Alice Wallenberg Foundation
through a Royal Swedish Academy of Sciences Fellowship, by the Swedish Research Council grants 642-2013-7837,  325-2009-7664, and
by the National Science Foundation under the CAREER Award DMR-0955902, A.~S. was supported by the Research Council of Norway, through
Grants 205591/V20 and 216700/F20. This work was also supported through the Norwegian consortium for high-performance computing (NOTUR).
\end{acknowledgments}

\bibliography{references}

\end{document}